\documentclass[onecolumn, a4size, 12pt]{IEEEtran}
\usepackage{amsmath}
\usepackage{amssymb}
\usepackage{amsfonts}
\usepackage{graphicx}
\usepackage{epsfig}
\usepackage{subfigure}
\usepackage{psfrag}

\title{Iterative Spectrum Shaping with Opportunistic Multiuser Detection
\footnote{R. Zhang is with the Institute for Infocomm Research,
A*STAR, Singapore (e-mail:rzhang@i2r.a-star.edu.sg).}\footnote{J. M.
Cioffi is with the Department of Electrical Engineering, Stanford
University, USA (e-mail:cioffi@stanford.edu).}}

\author{Rui Zhang and John M. Cioffi}

\begin{document}
\maketitle \vspace{-0.4in} \thispagestyle{empty}

\begin{abstract}
This paper studies a new decentralized resource allocation strategy,
named \emph{iterative spectrum shaping} (ISS), for the
multi-carrier-based multiuser communication system, where two
coexisting users independently and sequentially update transmit
power allocations over parallel subcarriers to maximize their
individual transmit rates. Unlike the conventional iterative
water-filling (IWF) algorithm that applies the single-user detection
(SD) at each user's receiver by treating the interference from the
other user as additional noise, the proposed ISS algorithm applies
multiuser detection techniques to decode both the desired user's and
interference user's messages if it is feasible, thus termed as
\emph{opportunistic multiuser detection} (OMD). Two encoding methods
are considered for ISS: One is \emph{carrier independent encoding}
where independent codewords are modulated by different subcarriers
for which different decoding methods can be applied; the other is
\emph{carrier joint encoding} where a single codeword is modulated
by all the subcarriers for which a single decoder is applied. For
each encoding method, this paper presents the associated optimal
user power and rate allocation strategy at each iteration of
transmit adaptation. It is shown that under many circumstances the
proposed ISS algorithm employing OMD is able to achieve substantial
throughput gains over the conventional IWF algorithm employing SD
for decentralized spectrum sharing. Applications of ISS in cognitive
radio communication systems are also discussed.
\end{abstract}


\setlength{\baselineskip}{1.3\baselineskip}
\newtheorem{claim}{Claim}
\newtheorem{guess}{Conjecture}
\newtheorem{definition}{Definition}
\newtheorem{fact}{Fact}
\newtheorem{assumption}{Assumption}
\newtheorem{theorem}{\underline{Theorem}}[section]
\newtheorem{lemma}{\underline{Lemma}}[section]
\newtheorem{corollary}{Corollary}
\newtheorem{proposition}{Proposition}
\newtheorem{example}{\underline{Example}}[section]
\newtheorem{remark}{\underline{Remark}}[section]
\newtheorem{algorithm}{\underline{Algorithm}}[section]
\newcommand{\mv}[1]{\mbox{\boldmath{$ #1 $}}}

\section{Introduction}

This paper is concerned with spectrum sharing in a multiuser
communication system based on multi-carrier modulation techniques
such as discrete multitone (DMT) for wired-line communication and
orthogonal frequency division multiplexing (OFDM) for wireless
communication. It is assumed that neither the users' transmitters
nor their receivers are collocated and as a result there is no
centralized control over the users' transmissions. In addition, all
users are assumed to transmit over the same frequency band and thus
possibly interfere with each other. The above scenario exists in
many wire-line/wireless broadband communication systems in practice,
e.g., the DMT-based digital subscriber line (DSL) network, and the
OFDM-based wireless {\it ad hoc} network.

The system of interest is in nature a competitive environment due to
the lack of cooperation among the users. Therefore, decentralized
strategies for allocation of users' transmit resources such as
powers, bit rates, bandwidths, and/or antenna beams become crucial
to the achievable system throughput. Consequently, a great deal of
valuable scholarly work has been done in the literature on this
study. For the conventional narrow-band spectrum sharing over
single-antenna slow-fading channels, distributed transmit power
control has been studied in, e.g., \cite{Foschini}--\cite{Sung}, for
minimizing the sum power consumption to meet with each individual
user's quality-of-service (QoS) requirement. Following the similar
problem formulation, decentralized joint power control and
beamforming have been studied in, e.g.,
\cite{Tassiulas}--\cite{Blum03b} for the case of multi-antenna
transceivers.  In \cite{Yu02}, a decentralized power allocation
strategy so-called {\it iterative water-filling} (IWF) was proposed
for a 2-user DSL system, where each of the two users independently
and sequentially updates transmit power levels over different
subcarriers so as to maximize individual transmit rate, subject to
the coexisting user's interference treated as additional background
noise at the receiver. Because of its practical advantages for
implementation, the IWF algorithm has been thoroughly investigated
in the subsequent literature. For example, in \cite{Chung03},
\cite{Tse05}, IWF has been studied for spectrum sharing scenarios
with more than two users. In \cite{Luo}, \cite{Palomar}, conditions
on the convergence of IWF have been rigourously characterized.
Motivated by IWF, semi-centralized and centralized power allocation
schemes for multiuser spectrum sharing have also been studied in
\cite{Huang}--\cite{Hung07} and \cite{Yu06a}--\cite{Yu06c},
respectively, all based on the primal-dual Lagrange duality
approach.

The existing works on decentralized/centralized resource allocation
schemes for multiuser spectrum sharing \cite{Foschini}--\cite{Yu06c}
have mostly assumed the {\it single-user detection} (SD) at the
receiver by treating the interference from the other coexisting
users as additional noise, mainly because of implementation ease of
the proposed schemes. During the past decade, multiuser detection
techniques (see, e.g., \cite{Verdu} and references therein) have
been thoroughly studied in the literature, and proved under many
circumstances to be able to provide substantial performance gains
such as rate improvement and decoding error reduction over the
conventional SD. This fact motivates this paper to make an attempt
to combine the well-known IWF with multiuser detection such that at
each iteration of user transmit adaptation, the corresponding user
is able to decode both the desired message and some/all of the
interference users' messages -- thereby reducing the overall
interference at the receiver -- if such decoding is feasible, thus
termed as \emph{opportunistic multiuser detection} (OMD). The
resultant new decentralized resource allocation algorithm is named
{\it iterative spectrum sharing} (ISS). Note that the proposed ISS
maintains the main advantage of IWF to be a purely decentralized
algorithm, while it improves over IWF via replacing the SD by the
more advanced OMD. With OMD, the transmission of the updating user
at each iteration subject to concurrent transmissions of the other
coexisting users can be generally modeled by the Gaussian
multiple-access channel (MAC) \cite{Cover}, whereas there is a key
difference pointed out as follows. Unlike the conventional MAC, the
coexisting users considered in this paper are non-cooperative in
allocating transmit rates/powers over subcarriers due to the lack of
centralized control over their transmissions. As a result, whether
OMD should be applied and over which subset of users it should be
applied depend on the instantaneous channel gains as well as the
interference users' power and rate allocations. Note that the OMD in
the context of this paper is analogous to the ``successive group
decoder (SGD)'' in the fading MAC with unknown channel state
information (CSI) at the user transmitters (see, e.g., \cite{Wang08}
and references therein). The main contributions of this paper are
summarized as follows:

\begin{itemize}
\item This paper considers two encoding methods for the proposed ISS. One is {\it carrier joint encoding} (CJE)
where a single codeword is modulated by all the subcarriers and is
decoded at the receiver by a single decoder. The other encoding
method is designed to maximally exploit the advantage of OMD, named
{\it carrier independent encoding} (CIE), where independent
codewords are modulated by different subcarriers and thus allow for
variable rate assignments and adaptive decoding methods. For both
encoding methods, this paper derives the optimal user power
allocation strategies to maximize individual transmit rate at each
iteration. The derived power allocation schemes are shown to be
non-trivial extensions of the standard ``water-filling'' (WF) power
control \cite{Cover} for IWF.

\item This paper investigates the converged user power spectrums by the proposed ISS,
and compares them to those by IWF for various system setups. Such
comparison reveals some important insights on why ISS is able to
outperform IWF in terms of the achievable system throughput for
decentralized spectrum sharing.
\end{itemize}

The rest of this paper is organized as follows. Section
\ref{sec:system model} presents the system model of
multi-carrier-based multiuser spectrum sharing. Section
\ref{sec:problem formulation} provides the problem formulations to
determine the optimal user power allocation policies for the
proposed ISS with CIE and CJE. Section \ref{sec:algorithms} presents
the solutions to the formulated problems. Section \ref{sec:numerical
results} provides the simulation results to demonstrate the
performance gains of ISS over IWF. Finally, Section
\ref{sec:conclusion} concludes the paper.

\section{System Model}\label{sec:system model}

Consider a typical spectrum sharing scenario where $K$ users
transmit independent messages to their corresponding receivers
simultaneously over the same frequency band. For the purpose of
exposition, in this paper it is assumed that $K=2$, while the
general case of $K>2$ is to be studied in the future work. Both the
users are assumed to adopt a multi-carrier (DMT/OFDM) -based
transmission and have the same symbol period and cyclic prefix (CP)
period that is assumed to be larger than the maximal signal
multipath spread of the two users. The total bandwidth for spectrum
sharing is equally divided into $N$ orthogonal sub-channels. For the
time being, it is assumed that perfect time and frequency
synchronization with reference to a common clock system have been
established for both the users prior to their data transmission. In
addition, it is assumed that the difference between the propagation
delays from the two user transmitters to either one of their
receivers is much smaller than the CP period and, thus, such delay
differences can be safely accommodated within the CP period.
Consider a block-based transmission for the two users with each
block consisting of $L$ DMT/OFDM symbols, while $L$ is usually a
large number to guarantee sufficient coding protection within each
block transmission. For typical wireless applications, it is also
assumed that the block duration is sufficiently small as compared to
the coherence time of any channel between the users. Thus, all the
channels involved in this paper can be assumed to be block fading
(BF), i.e., they are constant during each block transmission but can
vary from block to block. Based on the standard DMT/OFDM modulation
and demodulation, the discrete-time baseband signals for the system
of interest are given by
\begin{align}\label{eq:signal model}
y_{1,n}=\tilde{h}_{11,n}x_{1,n}+\tilde{h}_{21,n}x_{2,n}+z_{1,n}
\nonumber \\
y_{2,n}=\tilde{h}_{22,n}x_{2,n}+\tilde{h}_{12,n}x_{1,n}+z_{2,n}
\end{align}
where $n=1,\ldots,N$ is the subcarrier index; $x_{i,n}$ and
$y_{i,n}$ are the transmitted signal and received signal at
subcarrier $n$, respectively, for user $i=1,2$; $\tilde{h}_{11,n}$
and $\tilde{h}_{22,n}$ are the  ``direct'' channel complex
coefficients for user 1 and 2, respectively, at subcarrier $n$,
while $\tilde{h}_{21,n}$ and $\tilde{h}_{12,n}$ are the
``interference'' channel complex coefficients from user 2 to 1, and
from user 1 to 2, respectively, at subcarrier $n$; and $z_{i,n}$ is
the receiver noise at subcarrier $n$ for user $i=1,2$. Note that
both the block and symbol indexes are dropped in (\ref{eq:signal
model}) for conciseness. Without loss of generality, it is assumed
that $\{z_{i,n}\}, \forall i,n $ are independent circularly
symmetric complex Gaussian (CSCG) random variables (RVs) each having
zero mean and unit variance. It is also assumed that $x_{i,n}$'s are
independent RVs each with zero mean and respective variance
$p_{i,n}$, while $p_{i,n}$ denotes the transmit power allocated to
subcarrier $n$ of user $i$. Let $P_1$ and $P_2$ denote the average
transmit power constraint for user 1 and 2, respectively. It thus
holds that $\frac{1}{N}\sum_{n=1}^Np_{i,n}\leq P_i, i=1,2.$

Two encoding methods are considered at each user transmitter. One is
{\it carrier independent encoding} (CIE), where each subcarrier is
assigned an independent codebook and from each codebook a codeword
is chosen to be modulated into $L$ consecutive DMT/OFDM symbols at
the corresponding subcarrier in each block.  At the receiver, $N$
independent decoders are used to decode the corresponding messages
from different subcarriers. Let $r_{i,n}$ denote the rate of the
codebook assigned to user $i$ at subcarrier $n$. The average
transmit rate of user $i$ then becomes $R^{\rm
CIE}_i=\frac{1}{N}\sum_{n=1}^Nr_{i,n}$. The other encoding method is
{\it carrier joint encoding} (CJE), where a single codebook is used
for each block transmission and only one codeword is chosen from
this codebook and is modulated into all $N$ subcarriers of $L$
DMT/OFDM symbols. At the receiver, a single decoder is used to
decode the message from all the subcarriers. Let $R^{\rm CJE}_i$
denote the rate of this single codebook for user $i$. Comparing CIE
and CJE, it is easily seen that CIE requires more encoding and
decoding complexities over CJE, due to the use of independent
codebooks over different subcarriers. In addition, for the same
finite value of $L$, the effective codeword length for CIE is
reduced by a factor $1/N$ as compared to that for CJE, thus
resulting in inferior error-correcting capabilities. Therefore, the
existing multi-carrier-based transmission systems in practice have
all chosen to use CJE instead of CIE. Nevertheless, it is worth
noticing that CIE provides more flexibility over CJE in adaptive
rate assignments and decoding methods over subcarriers, which, as
will be shown later in this paper, can be a beneficial factor for
the proposed ISS under certain circumstances.

The system model considered in this paper is known as the {\it
2-user parallel Gaussian interference channel}, for which
characterization of the capacity region is in general still an
unsolved problem (see, e.g., \cite{Han} and references therein).
Nevertheless, achievable rates of this channel have been thoroughly
studied in the literature based on different assumptions on the
level of cooperations between the users for encoding and decoding as
well as power and rate allocations over the subcarriers. In this
work, we constrain our study on this channel by making the following
major assumptions:
\begin{itemize}
\item Each of the two users only has the knowledge on its own channel as well as
the channel from the other user's transmitter to its receiver.

\item Each of the two users {\it independently} and {\it sequentially} updates its transmit power allocations
over different subcarriers to maximize individual transmit rate.

\item Each of the two users is able to obtain the knowledge on transmit rates/rate (for CIE/CJE) of the
other user over subcarriers; and both the users employ the same type
of encoding method (CIE or CJE) and the same set of codebooks.
Thereby, at one user's receiver, it is possible to apply multiuser
detection (MD) to decode both the desired user's message and the
interference user's message.
\end{itemize}

Note that in the above assumptions, the first two are due to
practical considerations and are same as those made by the
conventional IWF proposed in \cite{Yu02},\footnote{More precisely,
in the first assumption on the known interference channel between
the users, only the channel gain is to be known for SD of IWF while
both the channel gain and phase information are required for MD of
the proposed scheme.} while the third assumption is a new one and is
not present in IWF where only the single-user detection (SD) is
applied. The decentralized resource allocation scheme motivated by
IWF while employing the more advanced MD is named {\it iterative
spectrum sharing} (ISS) in this paper.

\section{Problem Formulation} \label{sec:problem formulation}

In this section, problem formulations are provided for the users to
determine their transmit power and rate allocations over different
subcarriers at each iteration of transmit adaptation. Both encoding
methods, namely, CIE and CJE, are considered. For brevity, only user
1's transmit adaptation is addressed here, while the developed
results also apply to user 2.

Consider first CIE. At a particular iteration for user 1 to update
its transmission, since user 2's transmit powers $\{p_{2,n}\}$ and
rates $\{r_{2,n}\}$ over different subcarriers are fixed values, the
maximum transmit rate of user 1 at subcarrier $n$ with an arbitrary
allocated transmit power $p_{1,n}$ can be expressed as\footnote{For
the purpose of exposition, continuous rate and power values are
assumed in this paper. In addition, it is assumed that the optimal
Gaussian codebook is employed by the two users. The developed
results in this paper are readily extended to the more practical
cases with discrete power and rate values and/or non-optimal
modulation and coding schemes via, e.g., applying the optimal
discrete bit-loading algorithm with the ``SNR gap'' approximation
\cite{Cioffi}.}

\begin{equation}\label{eq:rate CIE}
r_{1,n}(p_{1,n})=\left\{
\begin{array}{ll} C(h_{11,n}p_{1,n}) & r_{2,n}\leq
C(\frac{h_{21,n}p_{2,n}}{1+h_{11,n}p_{1,n}})  \\
C(h_{11,n}p_{1,n}+h_{21,n}p_{2,n})-r_{2,n} &
C(\frac{h_{21,n}p_{2,n}}{1+h_{11,n}p_{1,n}})<r_{2,n}\leq
C(h_{21,n}p_{2,n})
\\ C(\frac{h_{11,n}p_{1,n}}{1+h_{21,n}p_{2,n}}) & r_{2,n}>
C(h_{21,n}p_{2,n})
\end{array} \right.
\end{equation}
where $C(x)\triangleq \log_2(1+x)$ is the capacity function of the
AWGN channel \cite{Cover}, while
$h_{11,n}\triangleq|\tilde{h}_{11,n}|^2$ and
$h_{21,n}\triangleq|\tilde{h}_{21,n}|^2$. The above result is
illustrated in the following three cases corresponding to the three
expressions of $r_{1,n}$ in (\ref{eq:rate CIE}) from top to bottom.
Note that the following discussions apply to any subcarrier $n$ of
user 1.

\begin{itemize}
\item {\it Strong Interference}: In this case, the received
interference signal power from user 2 at user 1's receiver is
sufficiently large such that the contained message with rate
$r_{2,n}$ can be first decoded by SD with user 1's signal taken as
additional Gaussian noise. After that, by reconstructing the
received user 2's signal and subtracting it from $y_{1,n}$, user 1's
message can be decoded by SD. The above operation is known as {\it
successive decoding} in the MAC \cite{Cover}.

\item {\it Moderate Interference}: In this case, the received signal
power from user 2 is not as large as that in the previous case of
strong interference and as a result, user 2's message can not be
directly decoded by SD. However, it is still feasible for user 1 to
apply {\it joint decoding} \cite{Cover} to decode both users'
messages.\footnote{Note that an alternative decoding method in this
case is successive decoding along with ``rate splitting''
\cite{Rimoldi} or ``time sharing'' \cite{Cover} encoding technique.
However, these techniques require certain cooperation between the
users and are thus not considered in this paper.} In this case, the
rate pair of the two users falls on the $45$-degree segment of the
corresponding MAC capacity region boundary \cite{Cover}.

\item {\it Weak Interference}: In this case, the received signal
power from user 2 is too weak to be decoded even without the
presence of user 1's signal. As such, user 1's receiver has the only
option of treating user 2's signal as the additional Gaussian noise
and applying SD to decode directly user 1's message. Note that the
above SD is used in the conventional IWF regardless of the received
signal power from the interference user (user 2).
\end{itemize}

From the above discussions, it is known that MD is applied in both
cases of strong and moderate interferences, but not in the case of
weak interference. Thus, user 1's receiver opportunistically applies
MD to the interference user signal if it has a sufficiently large
received power to be decoded either successively or jointly with the
desired user signal. Therefore, the MD in the context of this paper
is called {\it opportunistic multiuser detection} (OMD).

In Fig. \ref{fig:functions} (a), $r_{1}(p_1)$ in (\ref{eq:rate CIE})
is illustrated. For conciseness, the index $n$ is dropped here. It
is assumed that $p_2=1$, $r_2=0.5$, and $h_{21}=h_{11}=1$. Note that
in this case $r_2<C(h_{21}p_2)$ and thus OMD instead of SD should be
applied. The rate achievable by SD, denoted by $r^{\rm
SD}_1(p_1)=C(\frac{h_{11}p_{1}}{1+h_{21}p_{2}})$ from (\ref{eq:rate
CIE}), is also shown for comparison. It is observed that user 1's
rate with OMD is improved over that with SD, and $r_1(p_1)$ is the
minimum of the two functions defined as $f(p_1) \triangleq
C(h_{11}p_{1})$ and $h(p_1)\triangleq
C(h_{11}p_{1}+h_{21}p_{2})-r_{2}$, which are the rates achievable by
successive decoding and joint decoding, respectively. The threshold
value of $p_1$, denoted by $p_{th}$, for which $r_1(p_1)=f(p_1)$ if
$p_1\leq p_{th}$ and otherwise $r_1(p_1)=h(p_1)$, is obtained from
(\ref{eq:rate CIE}) as
\begin{equation}\label{eq:pth}
p_{th}=\frac{1}{h_{11}}\left(\frac{h_{21}p_{2}}{2^{r_{2}}-1}-1\right).
\end{equation}
Note that $p_{th}\geq 0$ if $r_2<C(h_{21}p_2)$.

With $r_{1,n}(p_{1,n})$ given in (\ref{eq:rate CIE}) for all $n$'s,
the problem can be formulated for user 1 to optimize its power and
rate allocations over subcarriers to maximize its average rate in
the case of CIE. This problem is denoted as (P1) and is expressed as
\begin{align}
\mbox{(P1)}~~\mathop{\mathtt{max}}_{p_{1,n}\geq 0, \forall n} & \
R^{\rm
CIE}_1(\{p_{1,n}\}):=\frac{1}{N}\sum_{n=1}^N r_{1,n}(p_{1,n}) \nonumber \\
\mathtt{s.t.} & \ \frac{1}{N}\sum_{n=1}^N p_{1,n}\leq P_1. \nonumber
\end{align}
After (P1) is solved, from the obtained solution for $p_{1,n}$ at
subcarrier $n$, the corresponding transmit rate and decoding method
can be obtained from (\ref{eq:rate CIE}). The solution of (P1) is
given later in Section \ref{subsec:CIE}.

Next, the case of CJE is considered. Recall that $R^{\rm CJE}_2$ and
$\{p_{2,n}\}$ are user 2's transmit rate value and power allocations
over subcarriers, respectively, which are all fixed for user 1's
transmit optimization. With joint encoding over all the subcarriers,
the maximum transmit rate of user 1 under arbitrary power
allocations $\{p_{1,n}\}$ is expressed as
\begin{equation}\label{eq:rate CJE}
R^{\rm CJE}_1(\{p_{1,n}\})=\left\{
\begin{array}{ll} \mathbb{E}[C(h_{11,n}p_{1,n})] & R^{\rm CJE}_2\leq
\mathbb{E}[C(\frac{h_{21,n}p_{2,n}}{1+h_{11,n}p_{1,n}})]  \\
\mathbb{E}[C(h_{11,n}p_{1,n}+h_{21,n}p_{2,n})]-R^{\rm CJE}_2 &
\mathbb{E}[C(\frac{h_{21,n}p_{2,n}}{1+h_{11,n}p_{1,n}})]<R^{\rm
CJE}_2 \leq \mathbb{E}[ C(h_{21,n}p_{2,n})]
\\ \mathbb{E}[C(\frac{h_{11,n}p_{1,n}}{1+h_{21,n}p_{2,n}})] & R^{\rm CJE}_2>
\mathbb{E}[C(h_{21,n}p_{2,n})]
\end{array} \right.
\end{equation}
where for notational brevity, $\mathbb{E}[\cdot]$ is used to
represent the operation $\frac{1}{N}\sum_{n=1}^N (\cdot)$. Note that
the rate $R^{\rm CJE}_1$ here is analogous to the ergodic capacity
in wireless fading channels where a sufficient long codeword spans
over all possible fading states and the codeword rate is the average
of all the instantaneous mutual information of the channel at
different fading states \cite{Goldsmith}. Similar to CIE, the three
rate expressions of $R^{\rm CJE}_1$ in (\ref{eq:rate CJE}) are also
achievable by successive decoding, joint decoding, and SD,
respectively, whereas there is a key difference that only one of
these decoding methods is applied over all the subcarriers for CJE,
in contrast to the case of CIE, where each subcarrier can be
independently assigned one of these decoding methods.\footnote{Due
to frequency-selective channel variation, it may be possible that at
some subcarriers of user 1, the interference channel gains from user
2 are sufficiently large such that if CIE is used, OMD can be
applied at these subcarriers to immediately remove the effect of
these interferences, while in the case of CJE, whether OMD can be
applied depends on the interference channel gains at all the
subcarriers.} Thus, unlike CIE, the user in the case of CJE does not
have the flexibility for transmit rate and decoding method
adaptations over different subcarriers, while it still can optimize
over transmit power allocations and choose the best decoding method
to maximize its transmit rate.

The problem for user 1 to optimize its power allocations in the case
of CJE is denoted as (P2), and is expressed as
\begin{align}
\mbox{(P2)}~~\mathop{\mathtt{max}}_{p_{1,n}\geq 0, \forall n} & \
R^{\rm
CJE}_1 (\{p_{1,n}\}) \nonumber \\
\mathtt{s.t.} & \ \frac{1}{N}\sum_{n=1}^Np_{1,n}\leq P_1. \nonumber
\end{align}
After solving the optimal power allocations in (P2), the maximum
transmit rate and its achievable decoding method can be obtained
from (\ref{eq:rate CJE}). The solution of (P2) is provided later in
Section \ref{subsec:CJE}.

\section{Optimal Power Allocation}\label{sec:algorithms}

In this section, (P1) and (P2) for the case of CIE and CJE,
respectively, are solved to obtain the optimal power allocations for
user 1 at each iteration of transmit adaptation. It is shown that
the obtained power allocation solutions in both cases are
non-trivial variations of the standard WF solution \cite{Cover},
which is employed in IWF.

\subsection{Carrier Independent Encoding}\label{subsec:CIE}

In this part, (P1) for the case of CIE is studied. The objective
function of (P1) is the sum of $N$ independent functions,
$r_{1,n}(p_{1,n})$'s, each of which can be easily shown to be a
concave function of $p_{1,n}$. Therefore, the objective function is
concave in $\{p_{1,n}\}$. In addition, the constraint of (P1) is a
linear function of $p_{1,n}$'s. Thus, (P1) is a convex optimization
problem, and thus can be solved via convex optimization techniques.

In (P1), the objective function is separable in $n$ while the
constraint is not. Therefore, the {\it Lagrange dual decomposition}
method, which has been applied in prior works (see, e.g.,
\cite{Huang}-\cite{Yu06c}), is also proposed here to decouple the
constraint in $n$, and thereby decomposes (P1) into a set of $N$
independent subproblems each for a different subcarrier. First, the
Lagrangian of (P1) is written as
\begin{align}\label{eq:Lagrangian}
\mathcal{L}(\{p_{1,n}\},\lambda)=\frac{1}{N}\sum_{n=1}^N
r_{1,n}(p_{1,n})-\lambda (\frac{1}{N}\sum_{n=1}^N p_{1,n}-P_1)
\end{align}
where $\lambda$ is the non-negative dual variable associated with
the power constraint. Then, the Lagrange dual function of (P1) is
defined as
\begin{eqnarray}\label{eq:Lagrange dual}
g(\lambda)=\max_{p_{1,n}\geq 0, \forall n}\mathcal{L}(\{p_{1,n}\}).
\end{eqnarray}
The value of the dual function serves as an upper bound on the
optimal value of the original (primal) problem, denoted by $r^{*}$,
i.e., $r^*\leq g(\lambda)$ for any $\lambda\geq 0$. The dual problem
of (P1) is then defined as $\min_{\lambda\geq 0} g(\lambda)$. Let
the optimal value of the dual problem be denoted by $d^*$, which is
achievable by the optimal dual solution $\lambda^*$, i.e., $d^*=
g(\lambda^*)$. For a convex optimization problem with a strictly
feasible point, the Slater's condition \cite{Boyd} is satisfied and
thus the duality gap, $r^* - d^* \leq 0$, is indeed zero for (P1).
This result suggests that (P1) can be equivalently solved by first
maximizing its Lagrangian to obtain the dual function for some given
dual variable $\lambda$, and then solving the dual problem over
$\lambda\geq 0$.

Consider first the problem for maximizing the Lagrangian to obtain
the dual function $g(\lambda)$ for some given $\lambda$. It is
interesting to observe that $g(\lambda)$ can be rewritten as
\begin{eqnarray}\label{eq:dual function rewrite}
g(\lambda)= \frac{1}{N}\sum_{n=1}^N g_n(\lambda)+\lambda P_1
\end{eqnarray}
where
\begin{eqnarray}\label{eq:dual function per subcarrier}
g_n(\lambda)=\max_{p_{1,n}\geq 0} r_{1,n}(p_{1,n})-\lambda p_{1,n}
~~ n=1,\ldots,N.
\end{eqnarray}
By this way, $g(\lambda)$ can be obtained via solving a set of $N$
independent subproblems, each for a different subcarrier $n$. Note
that the maximization problems in (\ref{eq:dual function per
subcarrier}) at different $n$'s all have the same structure and thus
can be solved using the same computational routine. For conciseness,
the index $n$ is dropped in (\ref{eq:dual function per subcarrier})
and the resultant problem is re-expressed as
\begin{align}
\mbox{(P3)}~~\mathop{\mathtt{max}}_{p_1\geq 0} \
a(p_1):=r_1(p_1)-\lambda p_1 \nonumber
\end{align}
where $r_1(p_1)$ is given by (\ref{eq:rate CIE}) with the index $n$
dropped.

Solutions of (P3) for all the subcarriers can then be used to obtain
the dual function $g(\lambda)$ in (\ref{eq:Lagrange dual}) for any
given $\lambda$. Then, the dual function needs to be minimized over
$\lambda\geq 0$ in the dual problem to obtain the optimal dual
solution $\lambda^*$ with which the duality gap is zero, i.e., the
original problem (P1) is equivalently solved. The standard routine
in convex optimization to iteratively update $\lambda$ toward its
optimal solution is via the bisection method \cite{Boyd} based on
the subgradient of $g(\lambda)$, which can be shown to be
$P_1-\frac{1}{N}\sum_{n=1}^N p_{1,n}$. When $\lambda=\lambda^*$, the
associated optimal solution of (P1), denoted by $\{p_{1,n}^*\}$,
satisfies $\frac{1}{N}\sum_{n=1}^N p_{1,n}^*=P_1$. For brevity, the
details of this standard routine are omitted here.

\begin{figure}
\centering{
 \epsfxsize=6.0in
    \leavevmode{\epsfbox{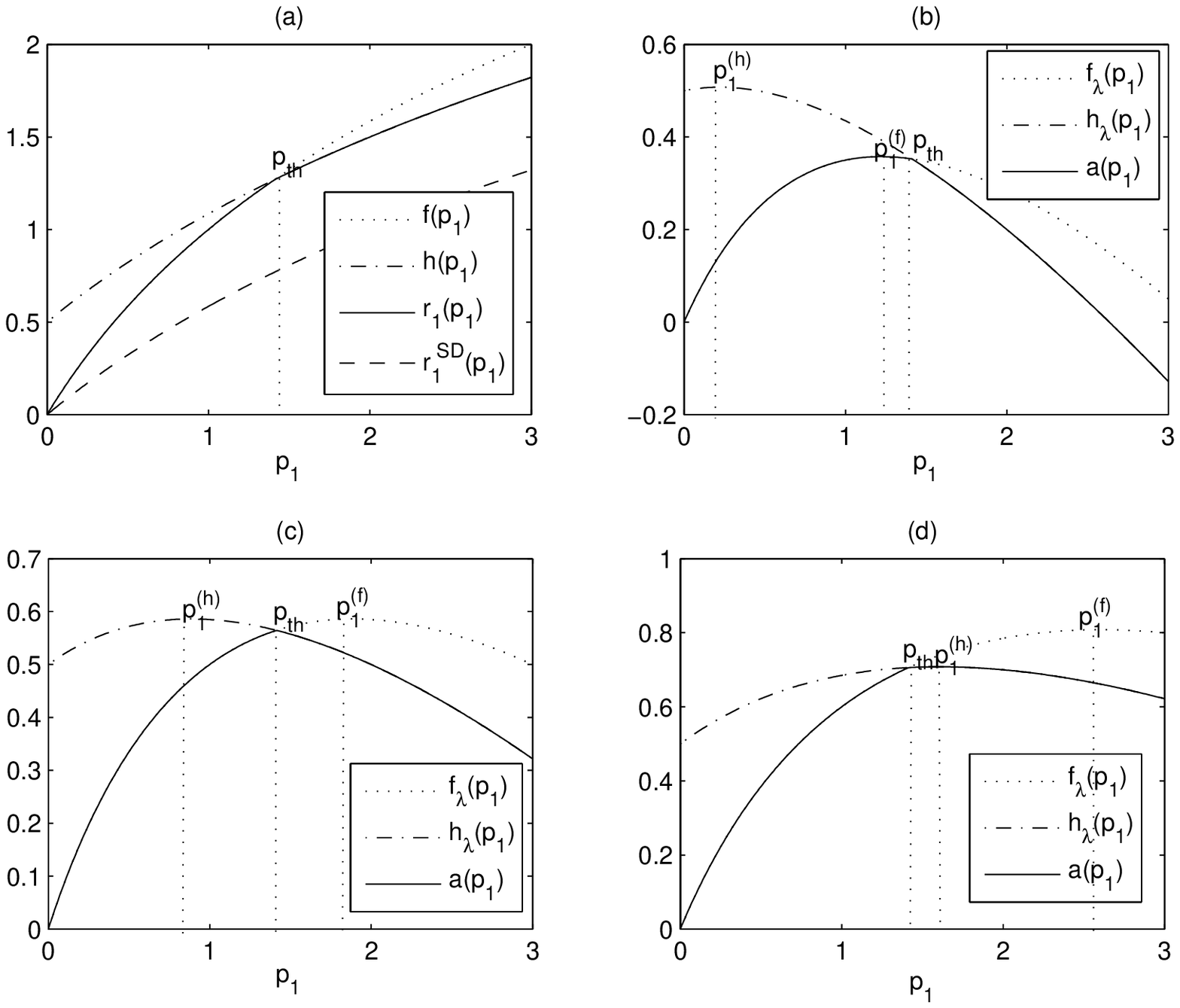}} }
\caption{Illustration of the functions $r_1(p_1)$ and
$a(p_1)\triangleq r_1(p_1)-\lambda p_1$ in the case of $r_2\leq
C(h_{21}p_2)$. Sub-figure (a) illustrates the function $r_1(p_1)$;
sub-figures (b), (c), and (d) illustrate the function
$a(p_1)=\min(f_{\lambda}(p_1),h_{\lambda}(p_1))$ for $\lambda=0.65,
0.5$, and $0.4$, respectively, where the function's maximum value is
achieved by $p_1^*=p_1^{(f)}, p_{th}$, and $p_1^{(h)}$,
respectively.}\label{fig:functions}
\end{figure}

Next, the solution of (P3) is derived for some given $\lambda$. Note
that since $r_1(p_1)$ is a concave function of $p_1$, so is $a(p_1)$
and thus (P3) is a convex optimization problem. The following
discussions are then made on the solution to (P3):

\underline{If $r_{2}\leq C(h_{21}p_{2})$}, from (\ref{eq:rate CIE})
it follows that OMD should be applied in this case. Note that
$p_{th}$ given in (\ref{eq:pth}) satisfies $p_{th}\geq 0$ in this
case, and $a(p_1)$ is the minimum of two functions defined as
$f_{\lambda}(p_1)\triangleq f(p_1)-\lambda p_1$ and
$h_{\lambda}(p_1)\triangleq h(p_1)-\lambda p_1$, where $f(p_1)$ and
$h(p_1)$ are defined earlier in Section \ref{sec:problem
formulation}. Also note that when $p_1\leq p_{th}$,
$a(p_1)=f_{\lambda}(p_1)$; otherwise, $a(p_1)=h_{\lambda}(p_1)$. The
optimal values of $p_1$ that maximize $f_{\lambda}(p_1)$ and
$h_{\lambda}(p_1)$ can be obtained as the standard WF solutions
\begin{equation}\label{eq:p f}
p_1^{(f)}=(\frac{1}{(\ln2)\lambda}-\frac{1}{h_{11}})^+
\end{equation}
with $(\cdot)^+\triangleq\max(0,\cdot)$ and
\begin{equation}\label{eq:p h}
p_1^{(h)}=(\frac{1}{(\ln2)\lambda}-\frac{1+h_{21}p_2}{h_{11}})^+
\end{equation}
respectively. Note that $0\leq p_1^{(h)}\leq p_1^{(f)}$. Let $a^*$
denote the optimal value of (P3), which is achievable by the optimal
solution $p_1^*$, i.e., $a^*=a(p_1^*)$. Since
$\max_{p_1}\min(f_{\lambda}(p_1),h_{\lambda}(p_1))\leq \min(
f_{\lambda}(p_1^{(f)}), h_{\lambda}(p_1^{(h)}))$, it follows that
$a^*\leq f_{\lambda}(p_1^{(f)})$ and $a^*\leq
h_{\lambda}(p_1^{(h)})$. Based on this result, $p_1^*$ is obtained
for the following three cases:
\begin{itemize}
\item $p_{th}\geq p_1^{(f)}$: In this case, $a(p_1^{(f)})=f_{\lambda}(p_1^{(f)})$, thus
it follows that $a^*\geq f_{\lambda}(p_1^{(f)})$. Since it has been
shown that $a^*\leq f_{\lambda}(p_1^{(f)})$, it follows that
$a^*=f_{\lambda}(p_1^{(f)})$ and $p_1^*=p_1^{(f)}$, as shown in Fig.
\ref{fig:functions} (b).  Note that successive decoding is optimal
in this case.

\item $p_{th}\leq p_1^{(h)}$: Similar to the first case, it can be shown that $a^*=h_{\lambda}(p_1^{(h)})$
and thus $p_1^*=p_1^{(h)}$, as shown in Fig. \ref{fig:functions}
(d). Note that joint decoding is optimal in this case.

\item $p_1^{(h)}<p_{th}<p_1^{(f)}$: Since
$p_{th}<p_1^{(f)}$, it follows that $f_{\lambda}(p_1)$ is an
increasing function for $p_1\leq p_{th}$. Moreover, since
$a(p_1)=f_{\lambda}(p_1)$, for $p_1\leq p_{th}$, it follows that
$f_{\lambda}(p_{th})\geq a(p_1)$ for any $p_1\leq p_{th}$.
Similarly, it can be shown that $h_{\lambda}(p_{th})\geq a(p_1)$ for
any $p_1\geq p_{th}$. Since
$h_{\lambda}(p_{th})=f_{\lambda}(p_{th})$, it concludes that
$p_1^*=p_{th}$, as shown in Fig. \ref{fig:functions} (c). In this
case, either successive decoding or joint decoding achieves the
optimum, while this paper adopts the former due to its more
implementation ease over the latter.

\end{itemize}

\underline{If $r_{2}> C(h_{21}p_{2})$}, SD should be used. Note that
$p_{th}<0$ in this case. It is easy to show that the optimal
solution $p_1^*$ of (P3) in this case is same as $p_1^{(h)}$ in
(\ref{eq:p h}) obtained earlier. Note that this WF-based power
allocation policy is also used in IWF.

By summarizing the above discussions, the following theorem is
obtained:

\begin{theorem}
The optimal solution of (P1) at subcarrier $n$, $n=1,\ldots,N$, is
(with the index $n$ dropped for conciseness)
\begin{eqnarray}\label{eq:optimal sol CIE}
p_1^*=\left\{\begin{array}{ll} p_1^{(f)}, & p_{th}\geq p_1^{(f)} \\
p_{th}, & p_1^{(h)}<p_{th}<p_1^{(f)} \\
p_1^{(h)}, & 0\leq p_{th}\leq p_1^{(h)}  \\
p_1^{(h)}, & p_{th}<0
\end{array} \right.
\end{eqnarray}
where $p_{th}$ is given in (\ref{eq:pth}), while $p_1^{(f)}$ and
$p_1^{(h)}$ are given in (\ref{eq:p f}) and (\ref{eq:p h}),
respectively, with $\lambda=\lambda^*$. The corresponding optimal
decoding methods at subcarrier $n$ are (from top to bottom)
successive decoding, successive decoding, joint decoding, and SD,
respectively.
\end{theorem}

\begin{figure}
\psfrag{a}{(a)}\psfrag{b}{(b)}\psfrag{c}{(c)}\psfrag{d}{$w^{(h)}$}\psfrag{e}{$w^{(f)}$}
\psfrag{f}{$\frac{1}{(\ln2)\lambda^*}$}\psfrag{g}{$p_{th}$}
\begin{center}
\scalebox{1.2}{\includegraphics*[73pt,604pt][362pt,761pt]{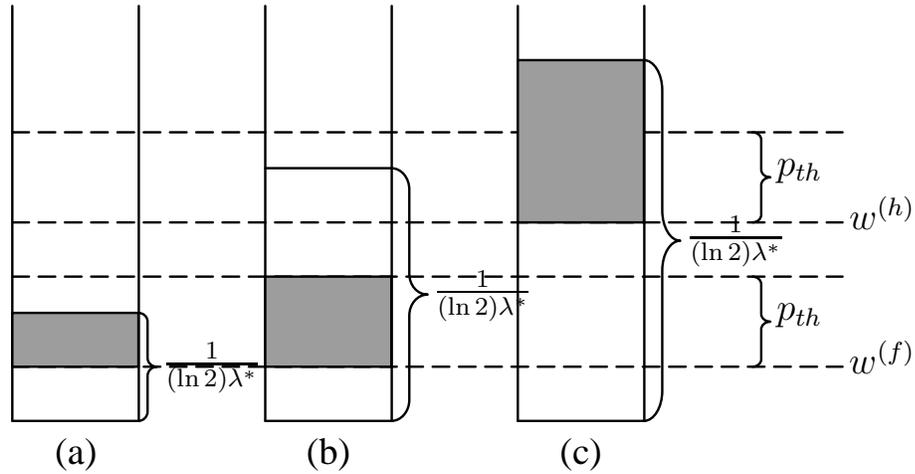}}
\end{center}
\caption{Illustration of the optimal power allocation
(\ref{eq:optimal sol CIE}) in the case of $r_2\leq C(h_1h_2)$: (a)
$w^{(f)}\leq \frac{1}{(\ln2)\lambda^*}\leq w^{(f)}+ p_{th}$; (b)
$w^{(f)}+p_{th}<\frac{1}{(\ln2)\lambda^*}<w^{(h)}+p_{th}$; and (c)
$\frac{1}{(\ln2)\lambda^*}\geq w^{(h)}+p_{th}$. The height of the
grey area in each case is the corresponding allocated
power.}\label{fig:WF}
\end{figure}

In Fig. \ref{fig:WF}, the optimal power allocation $p_1^*$ in
(\ref{eq:optimal sol CIE}) at a particular subcarrier $n$ is shown
for different values of $\lambda^*$. Note that $\lambda^*$ is a
decreasing function of user'1 average power constraint $P_1$. Only
the case of $r_2\leq C(h_{21}p_2)$ where OMD should be applied is
considered here. Thus, $p_{th}\geq 0$ and only the first three
expressions of $p_1^*$ in (\ref{eq:optimal sol CIE}) are illustrated
in this figure. It is observed that the obtained power allocation is
a variation of the standard WF solutions, e.g., $p_1^{(f)}$ in
(\ref{eq:p f}) and $p_1^{(h)}$ in (\ref{eq:p h}). There are two
fixed noise levels $w^{(f)}=1/h_{11}$ and
$w^{(h)}=(1+h_{21}p_2)/h_{11}$, corresponding to the power
allocations  $p_1^{(f)}$ and $p_1^{(h)}$, respectively. The amount
of power (water) to be allocated (filled) then depends on the
water-level $1/((\ln2)\lambda^*)$. If $P_1$ is sufficiently large
such that $1/((\ln2)\lambda^*)\geq w^{(f)}$ and at the same time
$P_1$ is sufficiently small such that $1/((\ln2)\lambda^*)\leq
w^{(f)}+p_{th}$, then $p_1^*=1/((\ln2)\lambda^*)-w^{(f)}=p_1^{(f)}$;
if $P_1$ is sufficiently large such that
$1/((\ln2)\lambda^*)>w^{(f)}+p_{th}$, but not yet large to make
$1/((\ln2)\lambda^*)\geq w^{(h)}+p_{th}$, then $p_1^*=p_{th}$
regardless of $\lambda^*$ and the resultant noise-plus-power level
is below the water-level $1/((\ln2)\lambda^*)$;\footnote{It is noted
that in realistic multi-carrier systems, the channel conditions vary
from subcarrier to subcarrier and as a result it is unlikely that
all the subcarriers will fall into this case and are thus allocated
powers $p_{th}(n)$'s regardless of $\lambda^*$ or $P_1$.} if $P_1$
is sufficiently large such that $1/((\ln2)\lambda^*)\geq w^{(f)}+
p_{th}$, then $p_1^*=1/((\ln2)\lambda^*)-w^{(h)}=p_1^{(h)}$. The
above three cases are illustrated by Fig. \ref{fig:WF} (a), (b), and
(c), respectively.

\subsection{Carrier Joint Encoding}\label{subsec:CJE}

Next, the problem (P2) for the case of CJE is studied. Similar to
the case of CIE, it can be shown that $R^{\rm CJE}_1(\{p_{1,n}\})$
in (\ref{eq:rate CJE}) is a concave function of $\{p_{1,n}\}$ and
thus (P2) is a convex optimization problem. Similar to (P1), the
Lagrange duality method is applied to solve (P2). Like (P1), the
Lagrangian and the dual function for (P2) can be obtained, and it
can be shown that (P2) has a zero duality gap. For brevity, these
details are skipped here and the min-max form of (P2) is directly
given as
\begin{equation}\label{eq:min max problem CJE}
\min_{\mu\geq 0} \max_{p_{1,n}\geq 0, \forall n} R^{\rm
CJE}_1(\{p_{1,n}\})-\mu(\frac{1}{N}\sum_{n=1}^N p_{1,n}-P_1)
\end{equation}
with $\mu$ denoting the non-negative dual variable associated with
the transmit power constraint. The optimal dual solution of $\mu$,
denoted by $\mu^*$, in the above minimization problem can be
similarly obtained by the bisection method as in (P1). In the
following, the maximization problem in (\ref{eq:min max problem
CJE}) over $\{p_{1,n}\}$ with some fixed $\mu$ is addressed, which
can first be simplified as (by removing the irrelevant constant
term)
\begin{equation}
\mbox{(P4)}~~\max_{p_{1,n}\geq 0, \forall n} b(\{p_{1,n}\}):=R^{\rm
CJE}_1(\{p_{1,n}\})-\mu\mathbb{E}[p_{1,n}]. \nonumber
\end{equation}

Similar to (P3), the following two cases are studied for (P4):

\underline{If $R^{\rm CJE}_2\leq \mathbb{E}[C(h_{21,n}p_{2,n})]$},
it is known from (\ref{eq:rate CJE}) that OMD should be used in this
case. Compared with the previously studied case of CIE, the power
optimization in the case of CJE is more involved, as explained as
follows: From (\ref{eq:rate CJE}), it is easy to show that if
$R^{\rm CJE}_2\leq \mathbb{E}[C(h_{21,n}p_{2,n})]$, $R^{\rm
CJE}_1(\{p_{1,n}\})$ can be expressed as the minimum of two
functions defined as $f_{\mu}(\{p_{1,n}\})\triangleq
\mathbb{E}[C(h_{11,n}p_{1,n})]-\mu\mathbb{E}[p_{1,n}]$ and
$h_{\mu}(\{p_{1,n}\})\triangleq
\mathbb{E}[C(h_{11,n}p_{1,n}+h_{21,n}p_{2,n})]-R^{\rm
CJE}_2-\mu\mathbb{E}[p_{1,n}]$. Then, let
$f_{\mu,n}(p_{1,n})\triangleq C(h_{11,n}p_{1,n})-\mu p_{1,n}$ and
$h_{\mu,n}(p_{1,n})\triangleq
C(h_{11,n}p_{1,n}+h_{21,n}p_{2,n})-R^{\rm CJE}_2-\mu p_{1,n}$,
$n=1,\ldots,N$, be the component in $f_{\mu}$ and $h_{\mu}$ at
subcarrier $n$, respectively, i.e., $f_{\mu}=\mathbb{E}[f_{\mu,n}]$,
$h_{\mu}=\mathbb{E}[h_{\mu,n}]$. Since $\min(f_{\mu},h_{\mu})$ is
not necessarily equal to $\mathbb{E}[\min(f_{\mu,n},h_{\mu,n})]$, it
is unclear whether $R^{\rm CJE}_1(\{p_{1,n}\})$ is separable in $n$,
which makes unclear whether the maximization of $b(\{p_{1,n}\})$
over $\{p_{1,n}\}$ is solvable directly by the dual decomposition
method.

Let $b^*$ denote the maximum value of $b(\{p_{1,n}\})$ achievable by
the optimal solution $\{p_{1,n}^*\}$. Note that both
$f_{\mu}(\{p_{1,n}\})$ and $h_{\mu}(\{p_{1,n}\})$ are concave
functions in $\{p_{1,n}\}$ and achieve their respective maximum
values at
\begin{align}
p_{1,n}^{(f)}&=(\frac{1}{(\ln2)\mu}-\frac{1}{h_{11,n}})^+, \ \
n=1,\ldots, N \label{eq:opt p CJE f} \\
p_{1,n}^{(h)}&=(\frac{1}{(\ln2)\mu}-\frac{1+h_{21,n}p_{2,n}}{h_{11,n}})^+,
 \ \ n=1,\ldots,N. \label{eq:opt p CJE h}
\end{align}
Note that $p_{1,n}^{(f)}\geq p_{1,n}^{(h)}, \forall n$. Since
$b(\{p_{1,n}\})=\min(f_{\mu}(\{p_{1,n}\}),h_{\mu}(\{p_{1,n}\})$, it
follows that $b^*\leq f_{\mu}(\{p_{1,n}^{(f)}\})$ and  $b^*\leq
h_{\mu}(\{p_{1,n}^{(h)}\})$. Next, the following cases are discussed
on $\{p_{1,n}^*\}$:
\begin{itemize}
\item $
\mathbb{E}[C(\frac{h_{21,n}p_{2,n}}{1+h_{11,n}p_{1,n}^{(f)}})]\geq
R^{\rm CJE}_2$: In this case,
$b(\{p_{1,n}^{(f)}\})=f_{\mu}(\{p_{1,n}^{(f)}\})$. Since $b^*\leq
f_{\mu}(\{p_{1,n}^{(f)}\})$, it follows that $b^*=
f_{\mu}(\{p_{1,n}^{(f)}\})$ and thus $p_{1,n}^*=p_{1,n}^{(f)}$. Note
that successive decoding is optimal in this case.

\item $\mathbb{E}[C(\frac{h_{21,n}p_{2,n}}{1+h_{11,n}p_{1,n}^{(h)}})]\leq R^{\rm
CJE}_2$: In this case,
$b(\{p_{1,n}^{(h)}\})=h_{\mu}(\{p_{1,n}^{(h)}\})$. Since $b^*\leq
h_{\mu}(\{p_{1,n}^{(h)}\})$, it follows that $b^*=
h_{\mu}(\{p_{1,n}^{(h)}\})$ and thus $p_{1,n}^*=p_{1,n}^{(h)}$.
Joint decoding is thus optimal.

\item $
\mathbb{E}[C(\frac{h_{21,n}p_{2,n}}{1+h_{11,n}p_{1,n}^{(f)}})]<R^{\rm
CJE}_2<\mathbb{E}[C(\frac{h_{21,n}p_{2,n}}{1+h_{11,n}p_{1,n}^{(h)}})]$:
In this case, $\{p_{1,n}^*\}$ is neither $\{p_{1,n}^{(f)}\}$ nor
$\{p_{1,n}^{(h)}\}$. Furthermore, by contradiction it can be shown
that
$\mathbb{E}[C(\frac{h_{21,n}p_{2,n}}{1+h_{11,n}p_{1,n}^*})]=R^{\rm
CJE}_2$ must hold in this case. Thus, $\{p_{1,n}^*\}$ can be
obtained by solving either one of the following two equivalent
problems:
\begin{align}
\mbox{(P5)}~~\mathop{\mathtt{max}}_{p_{1,n}\geq 0, \forall n} & \ \mathbb{E}[C(h_{11,n}p_{1,n})]-\mu\mathbb{E}[p_{1,n}] \nonumber \\
\mathtt{s.t.} & \
\mathbb{E}[C(\frac{h_{21,n}p_{2,n}}{1+h_{11,n}p_{1,n}})] \geq R^{\rm
CJE}_2. \nonumber \\
\mbox{(P6)} ~~ \mathop{\mathtt{max}}_{p_{1,n}\geq 0, \forall n} & \
\mathbb{E}[C(h_{11,n}p_{1,n}+h_{21,n}p_{2,n})]-R^{\rm CJE}_2
-\mu\mathbb{E}[p_{1,n}] \nonumber \\
\mathtt{s.t.} & \
\mathbb{E}[C(\frac{h_{21,n}p_{2,n}}{1+h_{11,n}p_{1,n}})]\leq R^{\rm
CJE}_2. \nonumber
\end{align}
Note that the objective functions of (P5) and (P6) are both concave
in $\{p_{1,n}\}$. However, (P5) is a non-convex optimization problem
since its constraint is not necessarily convex  due to the fact that
$C(\frac{b}{1+ax})$ is a convex function of $x$ for $x\geq 0$ with
any positive constants $a$ and $b$, while (P6) is a convex
optimization problem since its constraint has the reversed
inequality of that in (P5) and is thus a convex constraint.
Therefore, without loss of generality, (P6) is considered for this
case, while the obtained solution is optimal for both (P5) and (P6).
Similar to the third case of (P3), both successive decoding and
joint decoding achieve the maximum rate given the optimal power
allocations, whereas the former is more preferable than the latter
from an implementation viewpoint.
\begin{lemma}\label{lemma:new WF}
The optimal solution of (P6) is
\begin{equation}\label{eq:opt p new WF}
\tilde{p}_{1,n}^{(h)}=\left\{ \begin{array}{ll} 0, &
\frac{1}{(\ln2)\mu F_n(0)}-\frac{1+h_{21,n}p_{2,n}}{h_{11,n}}\leq 0
\\ x_n^*, &
\mbox{otherwise}
\end{array}\right.
\end{equation}
for $n=1,\ldots,N$, where $x_n^*$ is the unique positive root of the
equation
\begin{equation}\label{eq:x n}
x_n=\frac{1}{(\ln2)\mu F_n(x_n)}-\frac{1+h_{21,n}p_{2,n}}{h_{11,n}}
\end{equation}
while $F_n(x_n)$ is defined as
\begin{equation}\label{eq:F n}
F_n(x_n)=\frac{1+h_{11,n}x_n}{1+h_{11,n}x_n+\nu h_{21,n}p_{2,n}}
\end{equation}
and $\nu>0$ with which the constraint of (P6) is satisfied with
equality.
\end{lemma}
\begin{proof}
Please see Appendix \ref{appendix:proof lemma 2}.
\end{proof}
\end{itemize}
It is observed from (\ref{eq:opt p new WF}) and (\ref{eq:x n}) that
the optimal solution of (P6) resembles a {\it biased} version of the
standard WF solution $\{p_{1,n}^{(h)}\}$ given in (\ref{eq:opt p CJE
h}) because the associated water-level is biased by an additional
factor $F_n$, which itself is a function of the optimal power
allocation. It is also observed from (\ref{eq:F n}) that the biasing
factor is an increasing function of the allocated power. The
algorithm that resolves the biasing factor $F_n(x_n)$ to obtain the
solution of $x_n$ in (\ref{eq:x n}) is given in Appendix
\ref{appendix:algorithm}.

\underline{If $R^{\rm CJE}_2> \mathbb{E}[C(h_{21,n}p_{2,n})]$}, from
(\ref{eq:rate CJE}) it is known that SD should be applied at user
1's receiver in this case and $R^{\rm
CJE}_1(\{p_{1,n}\})=\mathbb{E}[C(\frac{h_{11,n}p_{1,n}}{1+h_{21,n}p_{2,n}})]$,
which is separable in $n$. Thus, $b(\{p_{1,n}\})$ is also separable
in $n$ and can be maximized independently over different $n$'s. It
is not hard to show that the optimal power allocations
$\{p_{1,n}^*\}$ in this case are equal to  $\{p_{1,n}^{(h)}\}$ given
in (\ref{eq:opt p CJE h}). Note that the power allocation policy
(\ref{eq:opt p CJE h}) is same as (\ref{eq:p h}), which is used in
IWF. Also note that the achievable rate of IWF is same with CIE or
CJE.

Summarizing the discussions on the above two cases, the following
theorem is obtained:
\begin{theorem}
The optimal solution of (P2) is
\begin{eqnarray}
p_{1,n}^*=\left\{\begin{array}{ll} p_{1,n}^{(f)}, &
\mathbb{E}[C(\frac{h_{21,n}p_{2,n}}{1+h_{11,n}p_{1,n}^{(f)}})]\geq
R^{\rm CJE}_2 \\ \tilde{p}_{1,n}^{(h)} , &
\mathbb{E}[C(\frac{h_{21,n}p_{2,n}}{1+h_{11,n}p_{1,n}^{(f)}})]<R^{\rm
CJE}_2<\mathbb{E}[C(\frac{h_{21,n}p_{2,n}}{1+h_{11,n}p_{1,n}^{(h)}})] \\
p_{1,n}^{(h)}, &
\mathbb{E}[C(\frac{h_{21,n}p_{2,n}}{1+h_{11,n}p_{1,n}^{(h)}})]\leq
R^{\rm
CJE}_2\leq \mathbb{E}[C(h_{21,n}p_{2,n})]  \\
p_{1,n}^{(h)}, & R^{\rm CJE}_2> \mathbb{E}[C(h_{21,n}p_{2,n})]
\end{array} \right.
\end{eqnarray}
for $n=1,\ldots,N$, where $p_{1,n}^{(f)}$, $p_{1,n}^{(h)}$, and
$\tilde{p}_{1,n}^{(h)}$ are given in (\ref{eq:opt p CJE f}),
(\ref{eq:opt p CJE h}), and (\ref{eq:opt p new WF}), respectively,
with $\mu=\mu^*$. The corresponding optimal decoding methods are
(from top to bottom) successive decoding, successive decoding, joint
decoding, and SD, respectively.
\end{theorem}

\section{Simulation Results} \label{sec:numerical results}

In this section, the performance of the proposed ISS algorithm with
OMD is evaluated and compared to that of the conventional IWF
algorithm with SD. It is assumed that the multi-carrier system has
the number of subcarriers $N=64$ and the CP period is equal to $1/4$
of the symbol period. All the channels involved in the system,
including users' direct channels and interference channels, are
assumed to each have 16 independent, equal-power, multipath taps. In
addition, a symmetric channel model is assumed where the two users'
direct channels have the same average unit power, and the two
interference channels between users have the same average power
denoted by $\rho$, while $\rho$ may take different values in order
to investigate the effect of the interference between the two users
on their achievable rates. In total, 1000 independent channel
realizations are simulated over which each user's achievable average
rate is computed, while the rate loss due to the insertion of CP is
ignored. For each channel realization, the multipath taps of the
direct/interference channels are generated by independent CSCG RVs
with zero mean and equal variance. The ISS/IWF algorithm is then
implemented over each channel realization where the two users
iteratively update their power allocations until their rates both
get converged.

\begin{figure}
\centering{
 \epsfxsize=5.0in
    \leavevmode{\epsfbox{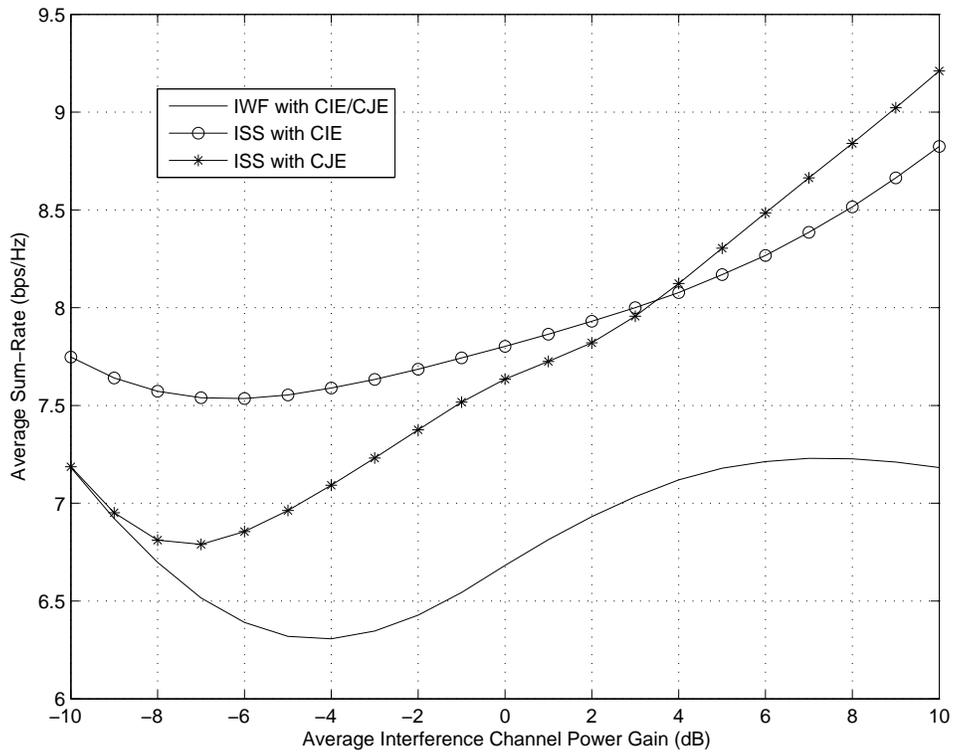}} }
\caption{The achievable sum-rate versus the average interference
channel power gain $\rho$ between the users for
$P_1=P_2=100$.}\label{fig:rate gain}
\end{figure}

\begin{figure}
\centering{
 \epsfxsize=5.0in
    \leavevmode{\epsfbox{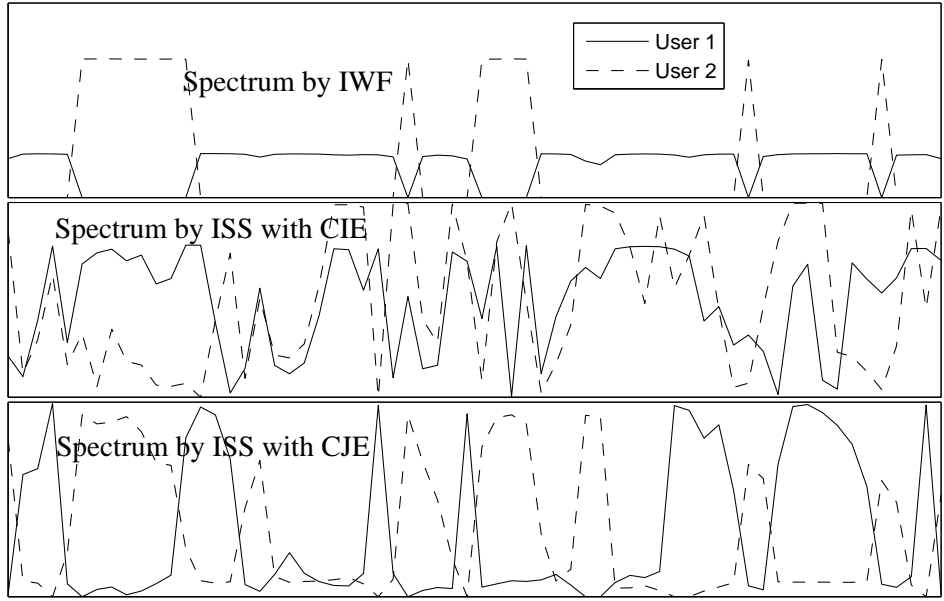}} }
\caption{A snapshot on the converged user power spectrums in the
case of $P_1=P_2=100$, and $\rho=10$.}\label{fig:spectrum}
\end{figure}

In Fig. \ref{fig:rate gain}, the achievable average sum-rate of the
two users is shown for different values of the interference channel
power gain, $\rho$. It is assumed that $P_1=P_2=100$. It is observed
that the proposed ISS algorithm with either CIE or CJE improves the
sum-rate over IWF, thanks to the more superior OMD over SD. It is
also observed that the achievable sum-rate of IWF fluctuates over
different values of $\rho$, while ISS ensures a consistent rate
increase with $\rho$ except the region of very low values of $\rho$
where OMD is not frequently applied. Interestingly, it is observed
that as $\rho$ increases, ISS with CJE becomes superior over that
with CIE in terms of the achievable sum-rate. Since CJE has a lower
complexity to implement than CIE, this result provides a useful
guidance for practical system design. However, this phenomenon is
some counter-intuitive since CIE provides each user more flexibility
for rate adaptations over different subcarriers and is thus expected
to be more suitable than CJE to exploit the benefit of OMD. A
reasonable explanation for this observation can be obtained by
looking at a snapshot of the users' converged power spectrums in
this case, as shown in Fig. \ref{fig:spectrum} for $\rho=10$. It is
observed that the two users' power spectrums in the case of IWF are
close to be orthogonal in frequency, which suggests that
``interference avoidance'' is probably the expected solution by IWF
in this case. In contrast, the power spectrums of the two users in
the case of ISS with CIE are observed to be almost overlapped in
frequency, as a result of OMD being applied at different
subcarriers, while the spectrums in the case of ISS with CJE appear
to be in between those of IWF and ISS with CIE. It is thus
conjectured that neither completely orthogonal nor overlapped
spectrum is the best converged solution for decentralized spectrum
sharing, which could probably explain why ISS with CJE performs the
best when the interference channel gains are large.

In Fig. \ref{fig:rate CR}, the achievable users' individual rates
are shown for a special case of the general channel model studied in
this paper. In this case, a ``cognitive radio'' type of newly
emerging wireless system is considered, where user 1 is the
so-called primary (non-cognitive) user (PU) that is the legitimate
user operating in the frequency band of interest, while user 2 is
the secondary (cognitive) user (SU) that transmits at the same time
over the same spectrum under the constraint that its transmission
will not cause the PU's QoS to an unacceptable level. Note that a
similar scenario has also been considered in \cite{Popovski07}. The
PU is non-cognitive since it is oblivious to the existence of the SU
and, thus, it applies the conventional IWF algorithm with SD by
treating the interference from the SU as additional noise. While for
the SU, it is cognitive in the sense that it is aware of the PU and
thus transmits with a much lower average power than that of the PU
in order to protect the PU. In this simulation, it is assumed that
$P_1=100$ and $P_2=1$. In addition, since the SU is cognitive, it
may choose to use the more advanced resource allocation scheme,
e.g., ISS with OMD instead of IWF with SD. Two cases are then
studied in this simulation: Case I, both user 1 and user 2 employ
IWF; Case II, user 1 employs IWF while user 2 employs ISS. Note that
in both cases, CJE is assumed for both users since the PU, with no
knowledge on the existence of the SU, should use CJE instead of CIE
from a practical consideration. In Fig. \ref{fig:rate CR}, it is
observed that the achievable rate of user 1 (the PU) drops slightly
in Case II as compared to Case I when $\rho$ is sufficiently large,
while the achievable rate of user 2 (the SU) improves significantly.
For example, at $\rho=1$, user 1's rate drop is only 3\% (a
negligible rate loss), while user 2's rate improvement is as large
as 140\% (a dramatic rate increase) by comparing Cases I and II.

\begin{figure}
\centering{
 \epsfxsize=5.0in
    \leavevmode{\epsfbox{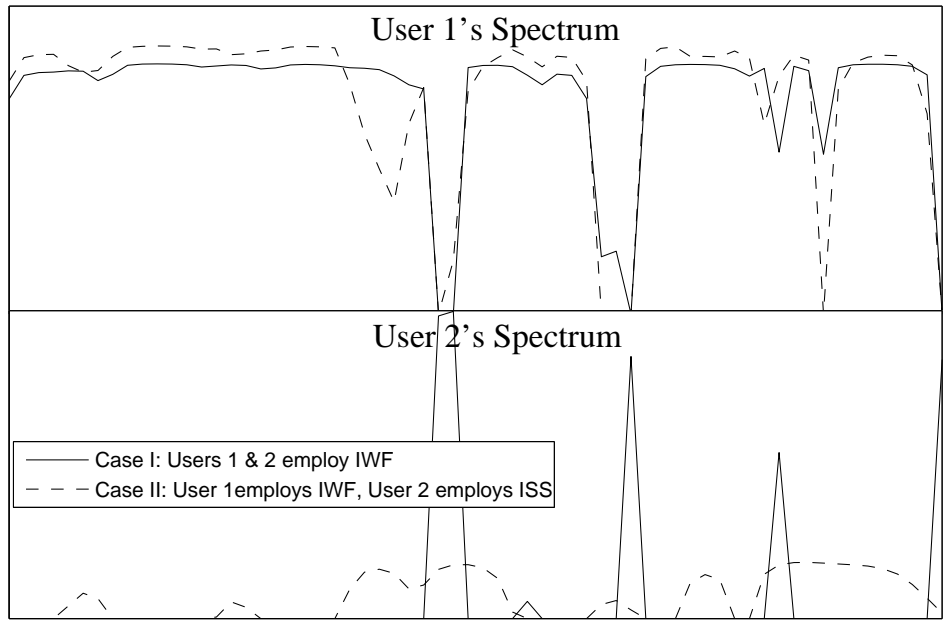}} }
\caption{A snapshot on the converged user power spectrums in the
case of $P_1=100$, $P_2=1$, and $\rho=5$ dB.}\label{fig:spectrum CR}
\end{figure}

The above observations can be explained by looking at a snapshot of
both users' converged power spectrums (normalized by users'
respective average powers) at a typical value of $\rho=5$ dB, as
shown in Fig. \ref{fig:spectrum CR}. It is observed that user 1's
spectrum does not change much over the two cases, while user 2's
spectrum changes dramatically from a very ``peaky'' one in Case I to
a more spread one in Case II. The SU's rate improvement in Case II
over Case I is due to OMD, which removes the effect of the PU's
interference and thus the SU can allocate powers based on its own
channel condition, while the PU's rate drop in Case II over Case I
is due to the ``interference diversity'' phenomenon
\cite{Zhang08TW}, namely, the more peaky interference in Case I is
more advantageous for minimizing the resultant PU's rate loss as
compared to the more spread one in Case II.

\section{Concluding Remarks} \label{sec:conclusion}

This paper studies a new decentralized resource allocation scheme,
ISS, for multi-carrier-based multiuser spectrum sharing. ISS
maintains the main advantages of the well-known IWF algorithm, e.g.,
being purely distributed and requiring only practical channel
knowledge, while it improves over IWF by exploiting OMD at the user
receiver. The resultant benefits are twofold: First, OMD improves
the user transmit rate at each iteration of resource adaptation as
compared to SD; Second, ISS with OMD leads to more balanced
converged user power spectrums than IWF with SD.

This paper presents the very initial results on ISS, for which many
issues remain unaddressed yet and are worth further investigating.
First, it is shown by simulation that for ISS, CJE performs better
than CIE with large interference channel gains, while the opposite
is true for moderate or small interference channel gains. This
observation raises the question on whether there exists an optimal
{\it multi-band encoding} scheme that divides the total bandwidth
into multiple sub-bands over which CIE is applied while within each
sub-band CJE is applied. Second, simulation results verify that the
convergence of ISS, like IWF, is always guaranteed with realistic
channel realizations, while characterizing the exact conditions for
the convergence of ISS is an important topic for the future study.
Last, extending the results of this paper to the cases with more
than two users and/or multi-antenna terminals will also be
interesting.

\appendices

\section{Proof of Lemma \ref{lemma:new WF}}\label{appendix:proof lemma 2}
Since (P6) is a convex optimization problem, the Lagrange dual
decomposition method can be applied to solve it, similar to that for
(P1). Let $\nu$ be the dual variable associated with the constraint
of (P6). Since it is already known that for the problem of interest
the constraint is satisfied with equality, it follows that $\nu>0$
from the Karush-Kuhn-Tucker (KKT) optimality condition \cite{Boyd}.
Then, (P6) can be written as the following equivalent min-max
optimization problem:
\begin{equation}
\min_{\nu>0}\max_{p_{1,n}\geq 0, \forall n }
\mathbb{E}[C(h_{11,n}p_{1,n}+h_{21,n}p_{2,n})]-R^{\rm CJE}_2
-\mu\mathbb{E}[p_{1,n}]-\nu(\mathbb{E}[C(\frac{h_{21,n}p_{2,n}}{1+h_{11,n}p_{1,n}})]-
R^{\rm CJE}_2)
\end{equation}
where the ``min'' part can be solved by the bisection method
similarly like (P1), while the ``max'' part for some given $\nu$ can
be decomposed into $N$ subproblems each for a different subcarrier.
At subcarrier $n$, the associated subproblem is expressed as
\begin{equation}
\max_{p_{1,n}\geq 0} C(h_{11,n}p_{1,n}+h_{21,n}p_{2,n}) -\mu
p_{1,n}-\nu C(\frac{h_{21,n}p_{2,n}}{1+h_{11,n}p_{1,n}})
\end{equation}
Let $\delta_n$ be the non-negative dual variable associated with the
constraint $p_{1,n}\geq 0$. The KKT optimality conditions for the
optimal primal and dual solutions of the above problem, denoted by
$p_{1,n}^*$ and $\delta_n^*$, respectively, are then obtained as
\begin{align}
p_{1,n}^*=\frac{1}{(\ln2)(\mu-\delta_n)
F_n(p_{1,n}^*)}-\frac{1+h_{21,n}p_{2,n}}{h_{11,n}},~
p_{1,n}^*\delta_n^*=0, ~p_{1,n}^* \geq 0, ~\delta_n^*\geq 0
\nonumber
\end{align}
where $F_n(\cdot)$ is given in (\ref{eq:F n}). From the above KKT
conditions, by considering the following two cases: (1)
$\delta_n^*>0$, $p_{1,n}^*=0$; and (2) $p_{1,n}^*>0$,
$\delta_n^*=0$,  (\ref{eq:opt p new WF}) can be correspondingly
obtained.

\section{Algorithm to Solve (\ref{eq:x n})} \label{appendix:algorithm}

\begin{figure}
\psfrag{a}{$G_n(x_n)$}\psfrag{b}{$x_n$}
\psfrag{c}{$\zeta_n$}\psfrag{d}{$\frac{1}{(\ln2)\mu}$}
\psfrag{e}{$x_n^*$}\psfrag{f}{$\measuredangle 45$}
\begin{center}
\scalebox{1.2}{\includegraphics*[50pt,525pt][290pt,750pt]{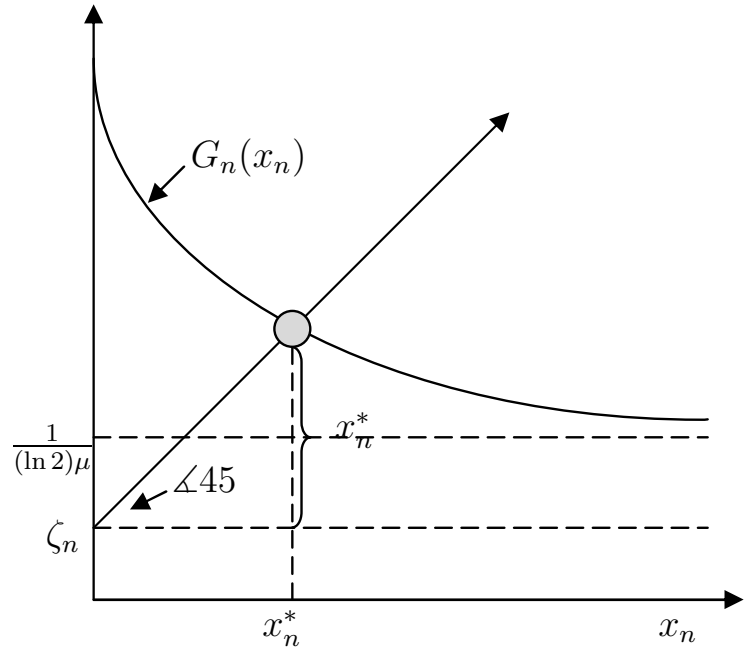}}
\end{center}
\caption{Illustration of the unique positive root $x_n^*$ for the
equation (\ref{eq:x n}).}\label{fig:Gn}
\end{figure}

The algorithm to obtain the unique positive root $x_n^*$ of the
equation (\ref{eq:x n}) is given in this appendix. Define
$G_n(x_n)=1/((\ln2)\mu F_n(x_n))$. Note that $G_n$ is a decreasing
function of $x_n$ for $x_n\geq 0$, and $G_n(0)\geq \zeta_n\triangleq
(1+h_{21,n}p_{2,n})/h_{11,n}$ from (\ref{eq:opt p new WF}), and
$G_n(\infty)=1/((\ln2)\mu)$. As shown in Fig. \ref{fig:Gn}, $x_n^*$
is then obtained as the intersection between a $45$-degree line
starting from the point $(0,\zeta_n)$ and the plot of the function
$G_n(x_n)$ in the region of $x_n\geq 0$. Numerically, $x_n^*$ can be
obtained by a simple iterative algorithm based on the bisection
search described as follows. Let $x_n^*\in[0,x_n^{\max}]$, where
$x_n^{\max}$ is an upper bound on $x_n^*$. A proper value of
$x_n^{\max}$ may be $G_n(0)-\zeta_n$ from Fig. \ref{fig:Gn}. For the
first iteration, let $\hat{x}_n$ be the midpoint of the initial
interval for $x_n^*$, i.e., $\hat{x}_n=\frac{1}{2}x^{\max}_n$. The
value of $G_n(\hat{x}_n)-\zeta_n$ is then computed, and compared to
$\hat{x}_n$: if it is larger than $\hat{x}_n$, it follows that
$x_n^* > \hat{x}_n$ and thus
$x_n^*\in(\frac{1}{2}x^{\max}_n,x^{\max}_n]$; otherwise, $x_n^*\leq
\hat{x}_n$ and $x_n^*\in[0,\frac{1}{2}x^{\max}_n]$. Thereby, after
the first iteration, the interval for searching $x_n^*$ is reduced
by half. The above process is repeated until $x_n^*$ is found within
any given accuracy.

\linespread{1.2}

\linespread{1.45}

\newpage

\begin{figure}
\centering{
 \epsfxsize=5.0in
    \leavevmode{\epsfbox{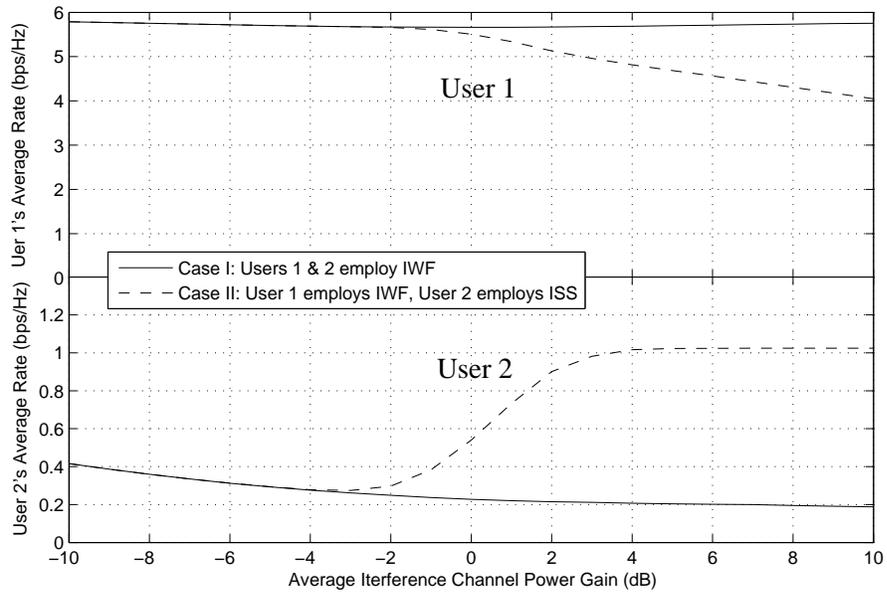}} }
\caption{The achievable user rates versus the average interference
channel power gain $\rho$ for $P_1=100$ and $P_2=1$.}\label{fig:rate
CR}
\end{figure}

\end{document}